# SPP-MIM hybridization meta-film: a biosensing structure uniting the merits of SPR and LSPR

Chenjia He, Xiaqing Sun, Tian Yang

*Abstract*—Having a flat device-solution interface is crucial for nanophotonic biosensors to achieve stable and reproducible performance, by mitigating solid-liquid-gas interfacial processes at the nanometer scale. In this aspect, the metal-insulator-metal (MIM) film presents a capable solution, by hybridizing surface plasmon polaritons (SPP) and MIM gap plasmons, which is enabled by the latter's unique dispersion characteristic and wide tunability. In this meta-film, the SPPs propagate along a flat interface, and seal the gap plasmons which can be integrated with nanostructures, e.g., a coupling grating. In addition, by tuning the gap plasmons, the SPP-MIM hybridization meta-film can be designed to achieve a significantly reduced SPP evanescent depth and a significantly improved surface sensitivity. Using gold as the plasmonic material, such improvements are theoretically predicted across a broad spectral range, from visible to near infrared. Particularly, at 1550 nm, we show that a grating-coupled meta-film device is designed to have its evanescent depth shortened from 1.4 μm to 0.16 μm, with an enhancement factor of 5.6 in its surface sensitivity, as compared with traditional grating-coupled SPR. This unique characteristic of the SPP-MIM meta-film makes it an efficient combination of propagating SPR and LSPR, by simultaneously having a flat and simple biosensing interface, a short evanescent depth and a high surface sensitivity. It provides an inspiring approach for transforming various LSPR designs into stable biosensors.

*Index Terms*—Surface plasmon resonance, localized surface plasmon resonance, metal-insulator-metal, hybridization.

## I. INTRODUCTION

Surface plasmon resonance (SPR) provides one of the most widely employed label-free biosensing technologies which detect the changes in surface refractive indices [1], [2]. The capability to measure biomolecular binding and dissociation kinetics in real time, with high precision and stability, has made SPR the current gold standard for biomolecular interaction analysis, with its applications increasingly widespread across various fields, including fundamental biology [3], health sciences [4], [5], [6], and drug development [7]. Further, it has long been expected that SPR sensing technologies would extend to even much larger scale applications that require easy operations and short turn-around-times, for example, point-of-care medical testing [8], [9], agriculture produce inspection [10], [11] and pollution monitoring [12], [13]. To fulfill this goal, extensive efforts have been made during the past two decades or so to study miniaturized SPR devices, which presented significantly reduced sizes, costs and operation complexities. Localized surface plasmon resonance (LSPR) is a crucial concept for constructing these miniaturized devices due to its nanoscale dimensions, flexible optical coupling schemes, high surface sensitivities, and wide optical tunabilities [14], [15], [16].

To date, pursuing ever higher sensitivities has become the predominant theme of nanoplasmonic biosensing research [17], [18], [19]. However, without an equally stable and reproducible measurement, the high sensitivities can't be converted to convincingly low limits-of-detection (LOD) [20]. Even if we can manage to exclude nonspecific binding processes and environmental interferences by measuring pure samples and keeping the solution parameters (such as temperature) constant, the baseline drifts and random signal jumps inherent to nanostructured surfaces could still completely overwhelm the tiny analyte binding signals [21], [22].

In a recent publication, we attribute the general instability of nanoplasmonic devices to the solid-liquid-gas interfaces of the nanostructured physical surfaces, and the complicated and not-well-understood interfacial processes therein [22]. In particular, we proposed and experimentally demonstrated the existence of surface gas nanobubbles at a nanopatterned gold surface [23], [24], and how it resulted in instability and surface pollution. Noticing that planar-surface biosensors, in general, could achieve over-ten-times better stability than those with nanostructured surfaces, in that same report, we proposed sealing the nanostructures under a flat plasmonic surface to eliminate the surface gas nanobubbles, and to achieve not only a high surface sensitivity, but also a high stability and reproducibility.

The proposed sensing structure, as mentioned above, is a hybridized meta-film which couples a nanopatterned metal-insulator-metal (MIM) waveguide with a flat surface plasmon polariton (SPP) surface. It unites the merits of both SPR and

This work was supported by the National Natural Science Foundation of China (grant No. 62375166, 61975253), the National Infrastructures for Translational Medicine (Shanghai) and the Lumieres (Xu Yuan) Biotechnology Company. Numerical simulation was supported by the Center for High Performance Computing of Shanghai Jiao Tong University. *(Corresponding author: Tian Yang.)*

Chenjia He, Xiaqing Sun and Tian Yang are with State Key Laboratory of Photonics and Communications, Key Laboratory for Thin Film and Microfabrication of the Ministry of Education, School of Electronic Information and Electrical Engineering, Shanghai Jiao Tong University, Shanghai 200240, China (e-mail: chenjiahe@sjtu.edu.cn, sunxiaqing@sjtu.edu.cn, tianyang@sjtu.edu.cn).

Xiaqing Sun is also with School of Material Science and Engineering, Shanghai Jiao Tong University, Shanghai 200240, China.

LSPR, including a flat and stable sensing surface, a high surface sensitivity (due to a shallow surface wave), a wide tunability, and a flexible optical coupling scheme. Interestingly, its surface sensitivity lowers as the surface refractive index increases, which has been considered a unique characteristic of LSPR.

In this report, we present a comprehensive and detailed modeling of the SPP-MIM hybridization meta-film, and its sensing performance as a grating-coupled device. We start by studying the hybridization mechanism at 850 nm optical wavelength, which includes an identification of the guided modes in the dispersion diagram, showing an anti-crossing between the SPP and the symmetric waveguide mode of MIM, and an analysis of the mode profiles before and after hybridization. In particular, the effect of tuning the insulator layer is scrutinized. Then, by embedding a periodic nanoslit grating in the MIM waveguide, we study the grating-coupled sensing performance of the meta-film, which includes its photonic crystal band diagram, the resonance modes and spectra, and its linear bulk sensitivities and nonlinear surface sensitivities. Next, we repeat our numerical calculations at 633 nm and 1550 nm, using gap dielectrics with refractive indices, $n_{gap}$, of 1.45 (silicon dioxide), 2.0 (silicon nitride) and around 3.5 (silicon), showing the wide tunability of the SPP-MIM meta-film and its superior performance across a broad wavelength range. Comparing with traditional grating-coupled SPR (GCSPR), thanks to the shortened SPP evanescent depth after hybridization, the surface sensitivities of the SPP-MIM meta-films are enhanced by a factor of 2.4 at 633 nm ($n_{gap}$=2.0), 3.9 at 850 nm ($n_{gap}$=2.0), and 5.6 at 1550 nm ($n_{gap}$=3.48), respectively, with a flat and non-patterned gold (Au) surface.

Notably, 1550 nm is the standard wavelength for modern fiber-optic telecommunication, which could take advantage of a huge set of mature and high-performance devices developed by the telecom industry. However, the SPP evanescent depth is up to around 1.4 µm at 1550 nm in traditional SPR (on a flat Au surface), which results in a surface sensitivity 2.8 and 1.6 times lower than that at 633 nm and 850 nm in the GCSPR configuration, respectively. This has made 1550 nm an unfavorable choice for biomolecular sensing. Here we show that, the SPP-MIM meta-film is able to shorten the evanescent field at 1550 nm to 0.16 µm, so as to achieve a surface sensitivity 2.0 and 3.5 times as high as that of traditional GCSPR at 633 nm and 850 nm, respectively. The prospect to turn this standard fiber-optic telecommunication wavelength into a favorable choice in biosensing applications could be inspiring for the optical biosensing industry.

## II. Hybridization of SPP and MIM

The SPP-MIM meta-film consists of a top metallic thin film, an insulator or dielectric gap layer, and a bottom metallic layer, while the sample solution provides a dielectric background (Fig. 1a). The figure schematically illustrates that the SPPs propagating along the flat interface between the solution background and the top metallic film, and the gap plasmons guided within the dielectric gap layer of the MIM waveguide couple with each other by tunnelling through the thin metal film, and forms a hybridized guided wave. To simplify the modeling of mode hybridization, the bottom metallic layer is assumed to be infinitely thick in this section. That is, we will study an Insulator-Metal-Insulator-Metal (IMIM) structure. Throughout this paper, the metal used in our model is Au, and the background is water.

To calculate the dispersion diagrams and eigenmode profiles of SPP, MIM and the hybridized modes, in this section, we will adopt the Drude model for the dielectric constant of Au and neglect its ohmic loss. The layer thicknesses of the IMIM in Fig. 1a have been optimized using a particle swarm algorithm to maximize its grating-coupled surface sensitivity near 850 nm, which will be discussed later. Consequently, the top Au film is 6 nm thick, and the gap layer is 80 nm thick with a refractive index, $n_{gap}$=2.0.

First, we plot the dispersion curve of the uncoupled SPP, which propagates along the interface between water and an infinitely thick Au layer. It is labeled as *SPP* in Fig. 1b. As shown, this curve starts close to the lightline of water, and converges to $\omega_{SP,water} = \omega_P(1 + \varepsilon_{water})^{-1/2}$ at large $k_x$ [25], where $\omega_P$ is the bulk plasmon frequency of Au, and $\varepsilon_{water}$ is the dielectric constant of water. The electric field profiles of *SPP* at three different wavevectors, $k_x$, are plotted in Fig. 1c, with each figure including an *I* layer (water) and an *M* layer (Au). Here the real parts of the complex electric fields, $E_z$ and $E_x$, are plotted, where $x$ is the guided wave's propagation direction.

Then, we plot the dispersion curves of the uncoupled MIM, by replacing the top Au film with an infinitely thick Au layer. There are two lowest-order gap plasmon modes of the MIM waveguide, one being symmetric with respect to the central *x-y* plane of the gap layer, labeled as *S*, the other one being antisymmetric, labeled as *A* [26]. Both MIM modes are transverse magnetic (TM) waves. As indicated by their dispersion curves, both modes approach the gap-Au interface SPP at large $k_x$, with their frequencies converging to $\omega_{SP,gap} = \omega_P(1 + \varepsilon_{gap})^{-1/2}$, where $\varepsilon_{gap}$ is the dielectric constant of the gap layer. The corresponding electric field profiles of the MIM waveguide modes are also plotted in Fig.1c, for the same set of $k_x$ values as for the *SPP*, which clearly show their respective spatial symmetries.

The *S* mode is particularly interesting, since it has an upward-lifting dispersion curve, which is widely tunable by varying the gap layer's thickness and refractive index as will be discussed in the next section. When the gap layer becomes thin enough, *S* will even become a negative refraction mode, operating at frequencies above $\omega_{SP,gap}$ [26], [27], [28]. Therefore, the dispersion curve of *S* can be flexibly tuned to cross that of *SPP*, and achieve SPP-MIM hybridization within a broad wavelength range. At the same time, the *A* mode's dispersion curve lies beneath the gap lightline and is unable to couple with the *SPP* efficiently. For similar reasons, it is difficult to achieve efficient coupling between a dielectric waveguide and the *SPP*, making the SPP-MIM meta-film a unique design for SPP engineering.

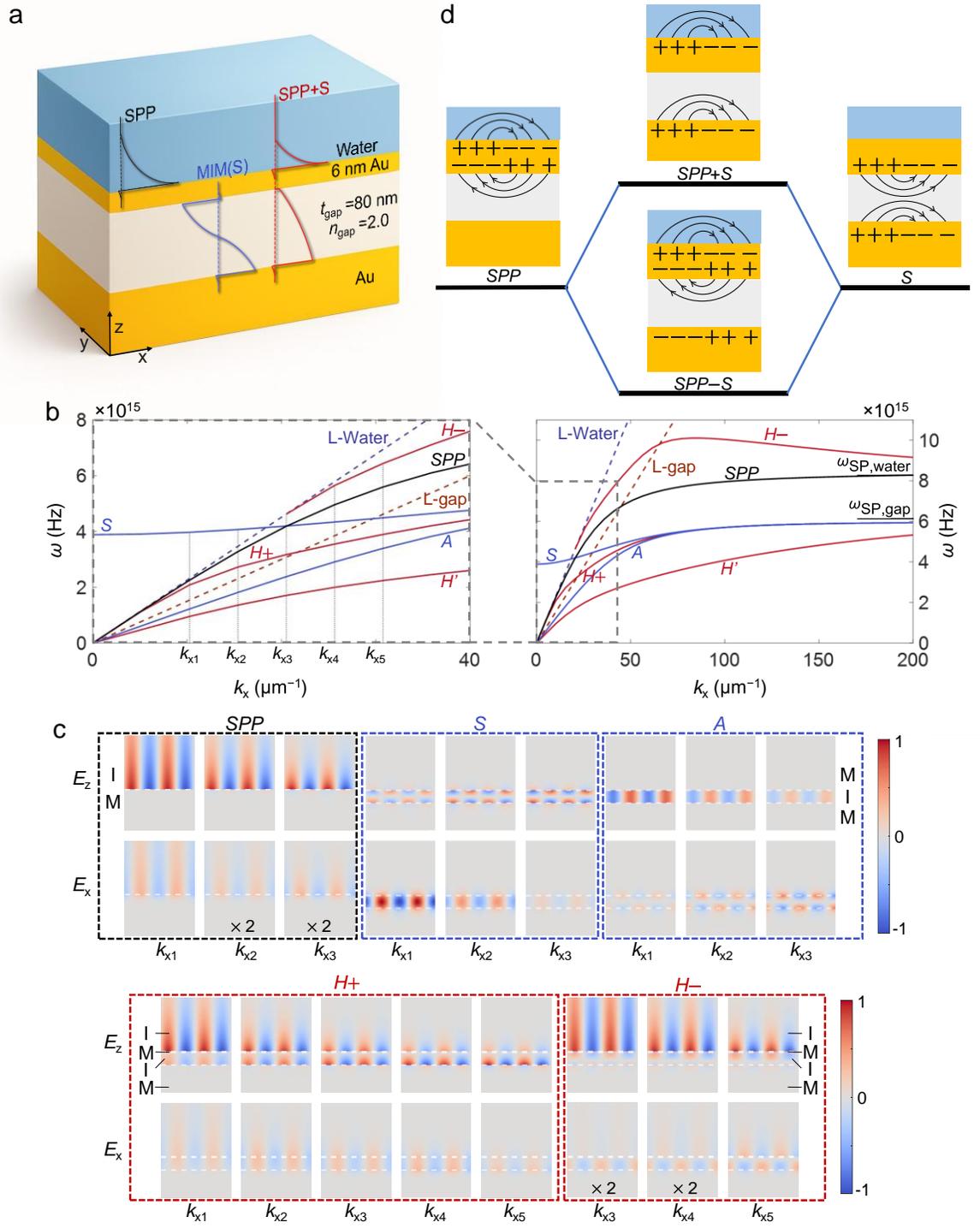

**Fig. 1. SPP-MIM hybridization.** (**a**) The IMIM structure. The SPP propagating along a water-Au interface, the MIM waveguide mode within the Au-dielectric-Au gap layer, and the hybridized mode are schematically illustrated. (**b**) The dispersion curves of *SPP*, MIM guided modes (*S* and *A*), and hybridized modes (*H−*, *H+* and *H'*). On the left is a zoom-in view of part of the dispersion diagram. (**c**) Mode profiles showing the real part of the complex *E*-field. Each subfigure spans two periods along the *x*-direction, and a total height of 640 nm along the *z*-direction. The *E*-fields are normalized to the maximum |**E**| for each mode. (**d**) Schematic illustration of the hybridization processes. The arrows indicate the *E*-fields, and the +/− symbols indicate the surface charges.

Lastly, the dispersion curves of the IMIM are also plotted in Fig. 1b, containing three modes labeled as *H'*, *H+* and *H−*, respectively. They approach the gap-Au SPP and the water-Au SPP at large $k_x$, respectively. At the crossing point of *S* and *SPP*, the frequencies and wavevectors of the two are both equal, resulting in a phase-matched coupling. This hybridization effect shows up as the anti-crossing between the dispersion curves of *H+* and *H−*. Note that *H−* disappears above the lightline of water at low frequencies, becoming a leaky mode there, while *H+* is a guided mode. In addition, the *A* mode mostly hybridizes with $H+$ to produce the $H'$ mode, which isn't interesting for biosensing applications and will not be discussed further.

The hybridization process is conceptually illustrated in Fig. 1d. When the linear superposition of $SPP$ and $S$ are in phase, we call it the $SPP+S$ mode, or the $H+$ mode. In this mode, the interference between $SPP$ and $S$ in the gap layer makes the $E_z$ component enhanced near the bottom of the gap but weakened near the top of the gap. Consequently, the electromagnetic energy of the $H+$ mode is localized both near the water-Au interface and the bottom of the gap. Such a mode profile enables an efficient excitation of the water-Au surface waves using bottom illumination, by embedding a grating into the bottom metal layer, as will be discussed later. The bottom illumination scheme is of crucial importance for fiber-tip integration, as demonstrated in our earlier work [21], [29]. The calculated field profiles of the $H+$ mode is plotted in Fig. 1c, at the same set of $k_x$ values as for plotting SPP and MIM, showing exactly what is predicted above. It is important to point out that, the dispersion curve of $H+$ is pushed away from that of $SPP$ by the hybridization effect, resulting in a remarkably larger $k_x$ (Fig. 1b) and a shallower evanescent surface wave at the water-Au interface (Fig. 1c), compared with that of $SPP$. This distinct characteristic of reduced evanescent depth endows the SPP-MIM meta-film the merits of both SPR and LSPR, and is the key to achieving a high surface sensitivity as will be shown later.

The hybridization process of the $SPP-S$ mode, or $H-$ mode is also shown in Fig. 1d, in which $SPP$ and $S$ superpose with a $\pi$ phase difference. Consequently, its electromagnetic energy is localized near the water-Au interface and the top of the gap. The calculated field profiles of $H-$ are plotted in Fig.1c. Here three larger $k_x$ values are used since $H-$ disappears at smaller $k_x$. The field profiles of $H+$ are also plotted at the same $k_x$ values for comparison purposes.

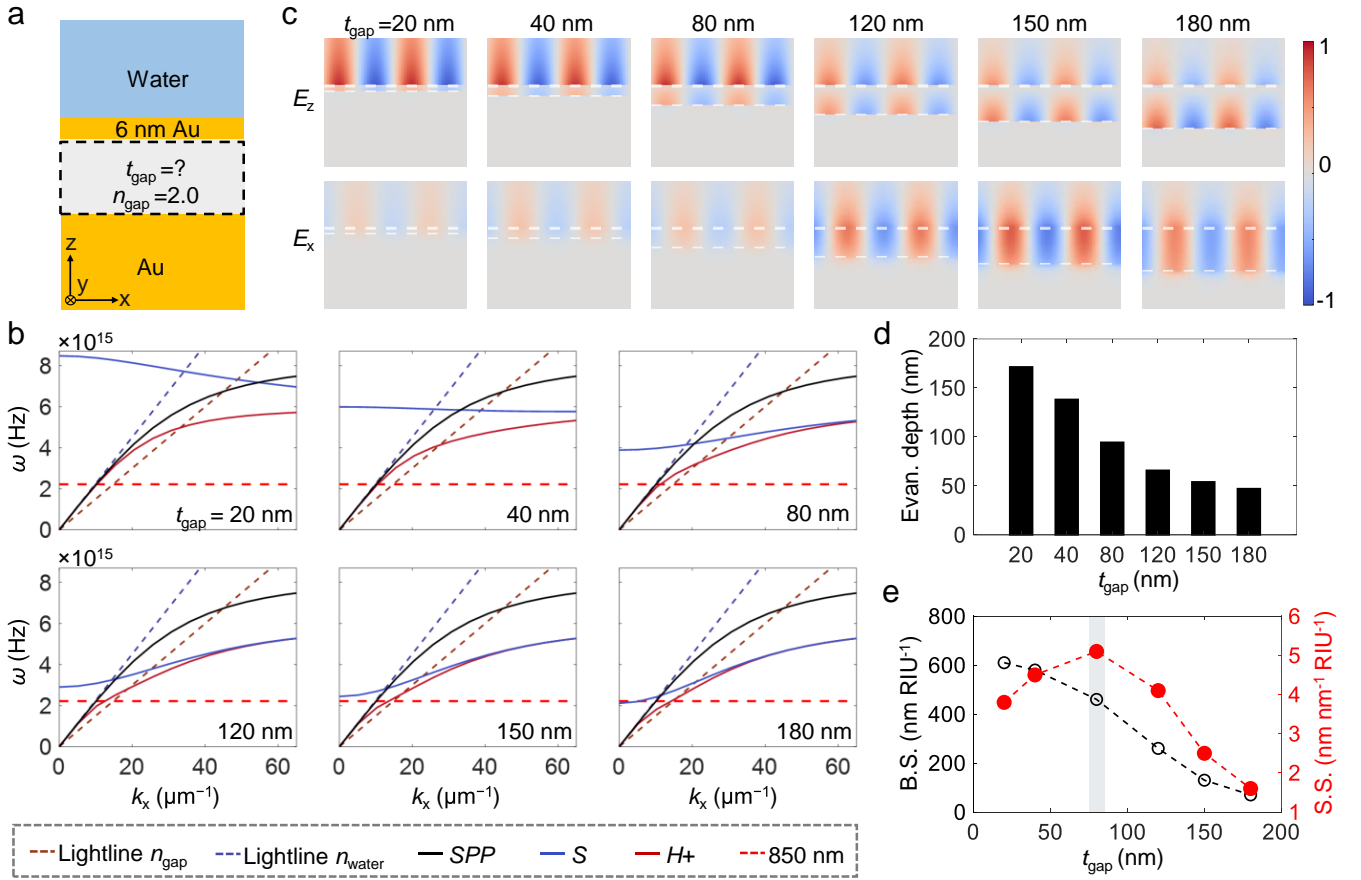

**Fig. 2. Tuning the dielectric gap thickness.** (a) The IMIM structure with a fixed gap refractive index, $n_{gap}$=2.0, and a variable gap thickness, $t_{gap}$. (b) The dispersion curves of $SPP$, $S$ and $SPP+S$ ($H+$) for different $t_{gap}$. (c) The $SPP+S$ mode profiles at 850 nm for different $t_{gap}$, showing the real part of the complex $E$-field. Each subfigure spans two periods along the $x$-direction, and a total height of 560 nm along the $z$-direction. The $E$-fields are normalized to the maximum $|E|$ for each case. (d) The evanescent depths of the $SPP+S$ mode with different $t_{gap}$. (e) The bulk sensitivities of the $SPP+S$ mode with different $t_{gap}$.

### III. TUNING THE HYBRID MODE – EFFECT OF THE GAP

In the remaining parts of this paper, we will focus on studying the $SPP+S$ hybrid mode, or the $H+$ mode. It was shown to have a high surface sensitivity and a low bulk susceptibility (due to the reduced evanescent depth of surface waves), a high biosensing stability (due to the flat physical surface), and a convenient optical grating-coupling scheme (due to electromagnetic energy localization near both the surface and the gap bottom) [22]. These distinct features of the *SPP+S* mode make the SPP-MIM meta-film a powerful solution for designing nanoplasmonic biosensing devices that are both sensitive and stable.

The dispersion curve of the *SPP+S* mode depends on where the *S* curve crosses the *SPP* curve (Fig. 1b). As mentioned earlier, the *S* curve is upward-lifted from the origin, that is, it has a non-zero frequency at $k_x$=0. In the following, we will show how to engineer and optimize this *SPP+S* hybridization by tuning the dielectric gap layer, including its effects on evanescent depth, bulk sensitivity, and surface sensitivity. In brief, increasing the gap thickness or refractive index results in a downward shift of the *S* dispersion curve and the *SPP+S* dispersion curve, which consequently reduces the evanescent depth of the latter in the water background, and a trade-off between the evanescent depth and the bulk sensitivity is needed to obtain the maximum surface sensitivity.

The same IMIM geometry as in the last section is considered, with the ohmic loss of Au neglected as well; the top Au film is also set to be 6 nm thick. First, we investigate the effect of the gap thickness, $t_{gap}$, with a fixed $n_{gap}$=2.0 (Fig. 2a). The dispersion curves of *SPP*, *S* and *SPP+S* are plotted in Fig. 2b for each value of $t_{gap}$, which clearly shows that the upward lifting of the *S* curve is reduced as $t_{gap}$ increases, due to a weakening in the spatial confinement of the *S* mode. The lowering of the *S* curve results in a simultaneous lowering of the *SPP+S* curve.

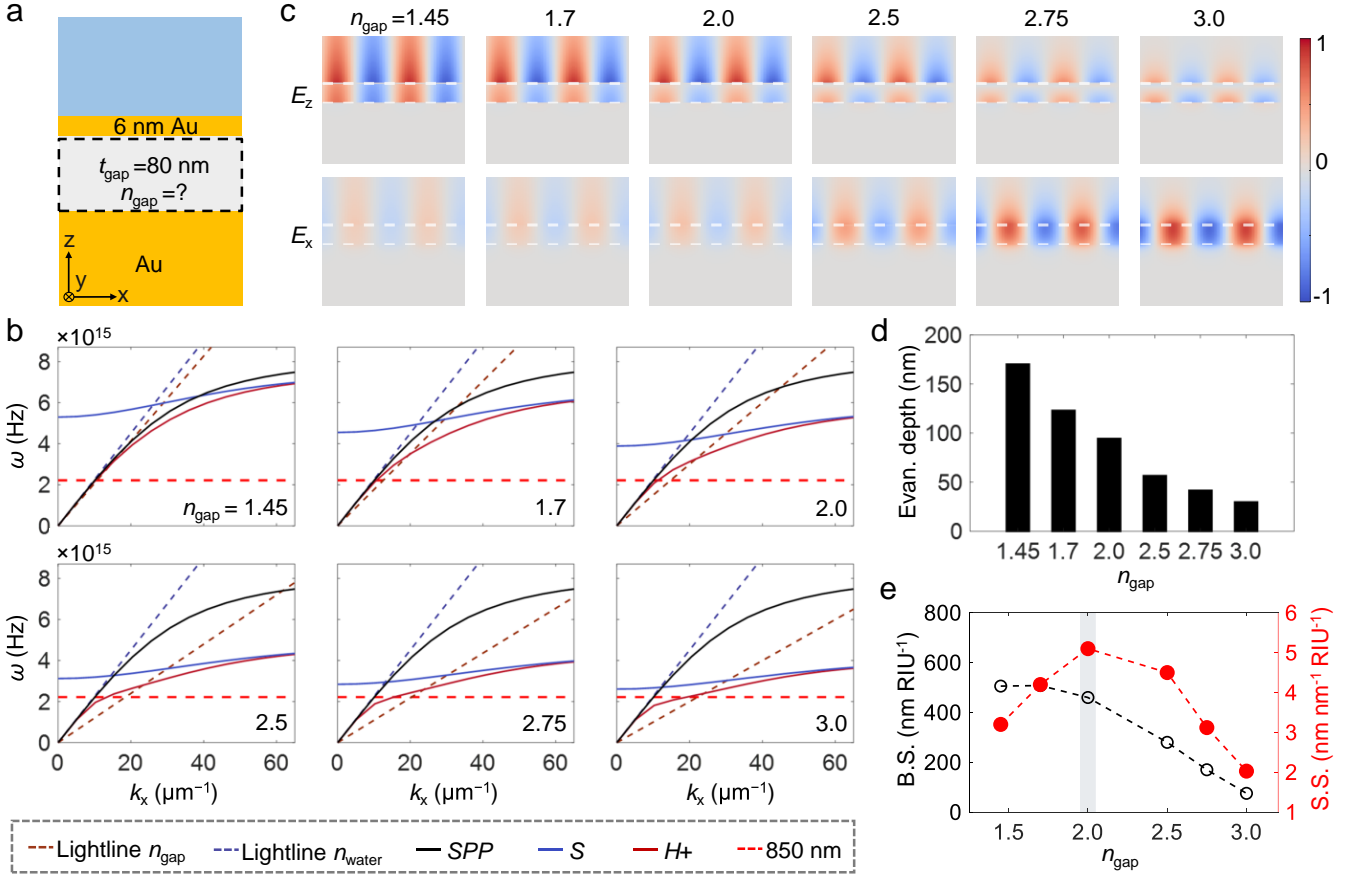

**Fig. 3. Tuning the dielectric gap's refractive index.** (**a**) The IMIM structure with a fixed gap thickness, $t_{gap}$=80 nm, and a variable gap refractive index, $n_{gap}$. (**b**) The dispersion curves of *SPP*, *S* and *SPP+S* (*H+*) for different $n_{gap}$. (**c**) The *SPP+S* mode profiles at 850 nm for different $n_{gap}$, showing the real part of the complex *E*-field. Each subfigure spans two periods along the *x*-direction, and a total height of 560 nm along the *z*-direction. The *E*-fields are normalized to the maximum |**E**| for each case. (**d**) The evanescent depths of the *SPP+S* mode with different $n_{gap}$. (**e**) The bulk sensitivities of the *SPP+S* mode with different $n_{gap}$.

If we aim to design a meta-film device operating at a predetermined optical wavelength, e.g., at 850 nm as marked by the red-dotted lines in Fig. 2b, the corresponding $k_x$ value of the *SPP+S* mode will increase from $1.03\times10^7$ m$^{-1}$ to $1.43\times10^7$ m$^{-1}$, as $t_{gap}$ increases from 20 nm to 180 nm. Meanwhile, the evanescent field in water follows $\exp(ik_x x - \frac{\alpha}{2} z)$, in which $k_x^2 - (\frac{\alpha}{2})^2 = k_0^2 n_{water}^2$. So the evanescent depth, $l_d = 1/\alpha$, will decrease from 172 nm to 48 nm, correspondingly. To show the trend of $l_d$ decreasing, the *E*-field profiles of the *SPP+S* mode at 850 nm are plotted in Fig. 2c, with the $1/e$-$|E^2|$ evanescent depths plotted in Fig. 2d. An important observation is that, the crossing point between the *SPP* curve and the *S* curve should be close enough to the target operation wavelength, in order to push the *SPP+S* curve away from the water lightline at the operation wavelength to achieve a remarkably shorter evanescent depth.

On the one hand, the shortening of evanescent depth favors detection of molecular binding at the device surface, and suppresses susceptibility to refractive index changes of the bulk background. This has been widely perceived as a unique merit of traditional LSPR biosensing as compared with SPR biosensing. On the other hand, as $t_{gap}$ increases, the field distribution of the $SPP+S$ mode transits from the meta-film's surface into its gap (Fig. 2c), which will ultimately diminish the surface sensitivity. To investigate the combined effect of these two factors, the grating-coupled bulk and surface sensitivities at each $t_{gap}$ are plotted in Fig. 2e. Obviously, a trade-off must be made between evanescent depth and bulk sensitivity to reach the maximum surface sensitivity. While the bulk sensitivity monotonically drops from 610 nm RIU$^{-1}$ at $t_{gap}$ = 20 nm to 70 nm RIU$^{-1}$ at $t_{gap}$ = 180 nm, the surface sensitivity peaks at $t_{gap}$ = 80 nm. In this section, the grating-coupled bulk and surface sensitivities are defined as the wavelength shift of the dispersion curve at a fixed $k_x$ (the grating's $k_x$) divided by the change of refractive index. The case of embedding a real grating will be discussed later. More details about sensitivity definitions and computation methods are included in *Appendix*.

Now let's investigate the effect of tuning the gap refractive index, $n_{gap}$. Here $t_{gap}$ is fixed at 80 nm, while $n_{gap}$ varies from 1.45 to 3.0 (Fig. 3a). The dispersion curves of $SPP$, $S$ and $SPP+S$ are plotted in Fig. 3b for each value of $n_{gap}$, which shows that the $S$ curve is lowered as $n$ increases, similar to traditional dielectric waveguides, and the $SPP+S$ curve is lowered simultaneously. The corresponding $E$-field profiles of the $SPP+S$ mode at 850 nm are plotted in Fig. 3c, with the $1/e$-|$E^2$| evanescent depths plotted in Fig. 3d. The grating-coupled bulk and surface sensitivities are plotted in Fig. 3e. As the $SPP+S$ dispersion curve is pushed away from the water lightline, the similar trends as tuning $t_{gap}$ are observed, including the shortening of evanescent depth, the decreasing of bulk sensitivity, and the trade-off to obtain a maximum surface sensitivity. Now we can conclude that, for the IMIM structure with a 6 nm top Au film, the maximum grating-coupled surface sensitivity of 5.1 nm nm$^{-1}$ RIU$^{-1}$ is obtained at $t_{gap}$ = 80 nm and $n$ = 2.0, where the evanescent depth is $l_d$ = 94 nm.

IV. NONLINEAR SURFACE SENSITIVITY

An intriguing characteristic of the SPP-MIM meta-film is that its surface sensitivity drops as more dielectric materials are deposited on its surface. This nonlinear behavior was previously perceived as a distinct characteristic of LSPR [30]. In Fig. 4, we perform an imaginary experiment, in which an $h$=10 nm thick dielectric film adheres to the surface of the meta-film, mimicking a layer of molecules, and the layer's refractive index, $n_s$, is increased in steps. The corresponding grating-coupled resonance wavelengths and evanescent depths are plotted. A continuous decrease of the surface sensitivity as $n_s$ increases can be observed, which drops from around 6.0 nm nm$^{-1}$ RIU$^{-1}$ at $n_s$ = 1.23 to around 2.9 nm nm$^{-1}$ RIU$^{-1}$ at $n_s$ = 1.73. Notably, before $n_s$ reaches 1.43, there is a fast increase of evanescent depth, which well explains the fast decrease of surface sensitivity in the same region.

The surface sensitivities elsewhere in the paper, unless specially mentioned, are for $n_s$ = 1.33.

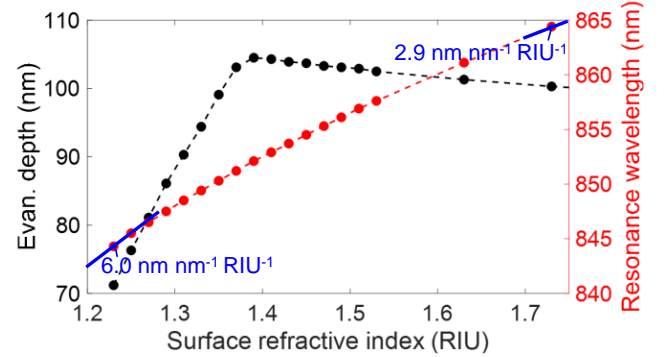

**Fig. 4. The surface sensitivity and evanescent depth changes with surface refractive index.** The slopes of the blue tangential lines indicate the surface sensitivities.

V. GRATING COUPLING AND PHOTONIC BANDS

Now let's embed a periodic nanoslit grating into the bottom Au layer of the meta-film to find out the device performance in a real grating-coupling configuration. Using a nanoslit grating to couple back illumination to SPP's has been previously studied for a single layer of metallic film, and employed for fiber end-facet biosensing [29], [31], [32]. Here we provide a theoretical analysis of the photonic band diagram and the mode field profiles for a nanoslit-grating-coupled SPP-MIM meta-film.

The device structure is schematically plotted in Fig. 5a. Here we follow the geometrical settings used in our previous work [22]. The bottom Au substrate in the IMIM is now replaced by a 65 nm thick Au film standing on a SiO$_2$ substrate. The top Au film is 18 nm thick. The gap has $n_{gap}$ = 1.45 (SiO$_2$) and $t_{gap}$ = 199 nm. A periodic array of nanoslits penetrate through the bottom Au film, with a period $\Lambda$ = 622 nm, and a slit width $w$ = 81 nm. Since this is a two-dimensional simulation, each nanoslit is infinitely long along the $y$ direction. To demonstrate the real performance that these grating-coupled meta-films can achieve, from now on, we will include the ohmic loss of Au in all of our calculations by using the Drude-Lorentzian model.

A TM planewave is set to be incident upon the device from the bottom at different angles (Fig. 5a). The power reflectivities for a range of wavelengths, $\lambda$, and $k_x$ are calculated to show the photonic band diagram of the device, which is effectively a one-dimensional photonic crystal (Fig. 5b). Similar to a single layer of metallic film [29], the 2$^{nd}$-order spatial Fourier component of the nanoslit array produces bandgaps at the center of the First Brillouin Zone. The bandedge states, labeled as points *I* to *VI*, have their corresponding |$E$|$^2$ field profiles shown in Fig. 5c. Note that for those dark modes at $k_x$ = 0, the field profiles are plotted at $k_x$ =0.08 μm$^{-1}$ or 0.18 μm$^{-1}$ instead.

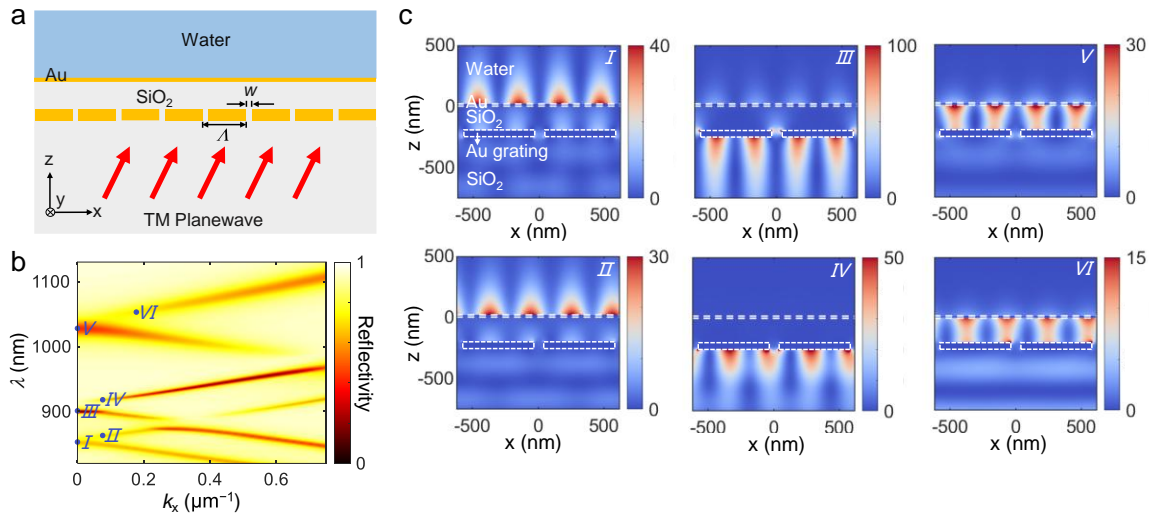

**Fig. 5. A meta-film embedded with a one-dimensional nanoslit photonic crystal.** (**a**) The simulation model (not to scale). (**b**) The simulated photonic band diagram, by plotting reflectivity versus ($\lambda$, $k_x$). (**c**) $|E|^2$ profiles for different bandedges, normalized to the incident light.

The bandedge points *I* and *II* belong to the *SPP+S* mode, with their electromagnetic energies localized near the water-Au interface and the gap's bottom (Fig. 5c). At $k_x = 0$, or equivalently, $\pm\frac{2\pi}{\Lambda}$, the grating mixes the two otherwise degenerate *SPP+S* propagating waves (without the grating) into the two standing-waves, *I* and *II*, with a $\frac{1}{4}\Lambda$ spatial shift to each other. Point *I* is a bright mode at $k_x = 0$, since it has a relatively high $|E_x|$ in the slits (note that for SPP, $E_x$ and $E_z$ has a $\pi/2$ phase difference), so that it couples with the normally incident illumination efficiently. At the same time, Point *II* is a dark mode at $k_x = 0$, since it has $|E_x| = 0$ at the center of each slit as indicated by the standing wave's symmetry, prohibiting coupling with the normally incident illumination.

The bandedge points *III* and *IV* belong to a different mode, which mainly comprises of SPP at the interface between the bottom Au film and the $SiO_2$ substrate (Fig. 5c). The points *V* and *VI* belong to the H' mode, which is mainly confined in the gap, and has been studied for sensing changes in the gap [33].

## VI. PERFORMANCE OF GRATING-COUPLED DEVICES AT VARIOUS WAVELENGTHS

In this section, we investigate what are the maximum surface sensitivities that the nanoslit-grating-coupled SPP-MIM meta-films can achieve, at 633, 850 and 1550 nm, respectively. Here, we focus on the bandedge point *I*, under back illumination at normal incidence, while it should be kept in mind that there are other possible configurations for exciting these meta-films. We will also only consider Au for the metal films, leaving discussions on using other plasmonic materials and the possible improvements of performance for the future.

To enhance the surface sensitivity, it is desirable to have a short evanescent depth. Therefore, researchers have been interested in the short wavelengths, including the visible and ultraviolet bands [34]. On the other hand, although 1550 nm is the standard wavelength for fiber-optic telecommunication, it has been an unfavorable choice for SPR biosensing due to its long evanescent depth and low surface sensitivity. We will show that, using the nanoslit-grating-coupled SPP-MIM meta-film, a remarkable enhancement of surface sensitivity can be achieved for a broad range of wavelengths. Particularly, a high surface sensitivity can be obtained at 1550 nm as well, making it a promising candidate for fiber-tip plasmonic sensing [35].

TABLE I
OPTIMIZED GEOMETRIC PARAMETERS OF NANOSLIT-GRATING-COUPLED DEVICES[a]

| $n_{gap}$ | $\lambda$ (nm) [b] | Geometric parameters (nm) | | | | |
|---|---|---|---|---|---|---|
| | | $t_{Au1}$ [c] | $t_{gap}$ | $t_{Au2}$ [d] | $\Lambda$ | $w_{slit}$ [e] |
| 3.48 | 1550 | 2 | 73 | 69 | 1010 | 85 [f] |
| 2 | 1550 | 5 | 168 | 47 | 1110 | 105 |
| 2 | 850 | 11 | 84 | 29 | 585 | 95 |
| 2 | 633 | 40 | 80 | 67 | 435 | 180 |
| 1.45 | 1550 | 5 | 178 | 22 | 1150 | 50 |
| 1.45 | 850 | 14 | 86 | 30 | 620 | 60 |
| 1.45 | 633 | 48 | 84 | 29 | 440 | 195 |
| / | 1550 | / | / | 40 | 1165 | 15 |
| / | 850 | / | / | 40 | 640 | 55 |
| / | 633 | / | / | 40 | 470 | 85 |

[a] Each of the top seven rows corresponds to a nanoslit-grating-coupled SPP-MIM meta-film with a set of optimized geometric parameters to obtain the maximum surface sensitivity at the desired resonant wavelength and the predetermined gap refractive index.
Each of the bottom three rows corresponds to a single Au layer, representing traditional GCSPR. Here, the $t_{Au2}$ values are taken to be large enough to have the water-Au interface SPR fairly decoupled from the $SiO_2$ substrate.
[b] The resonant wavelength (bandedge point *I*).
[c] The thickness of the top Au layer.
[d] The thickness of the bottom Au layer.
[e] The width of each nanoslit.
[f] In this row, $w_{slit}$ takes a larger value than what corresponds to the maximum surface sensitivity, to obtain a reasonably deep spectral resonance dip (Fig. 6b).

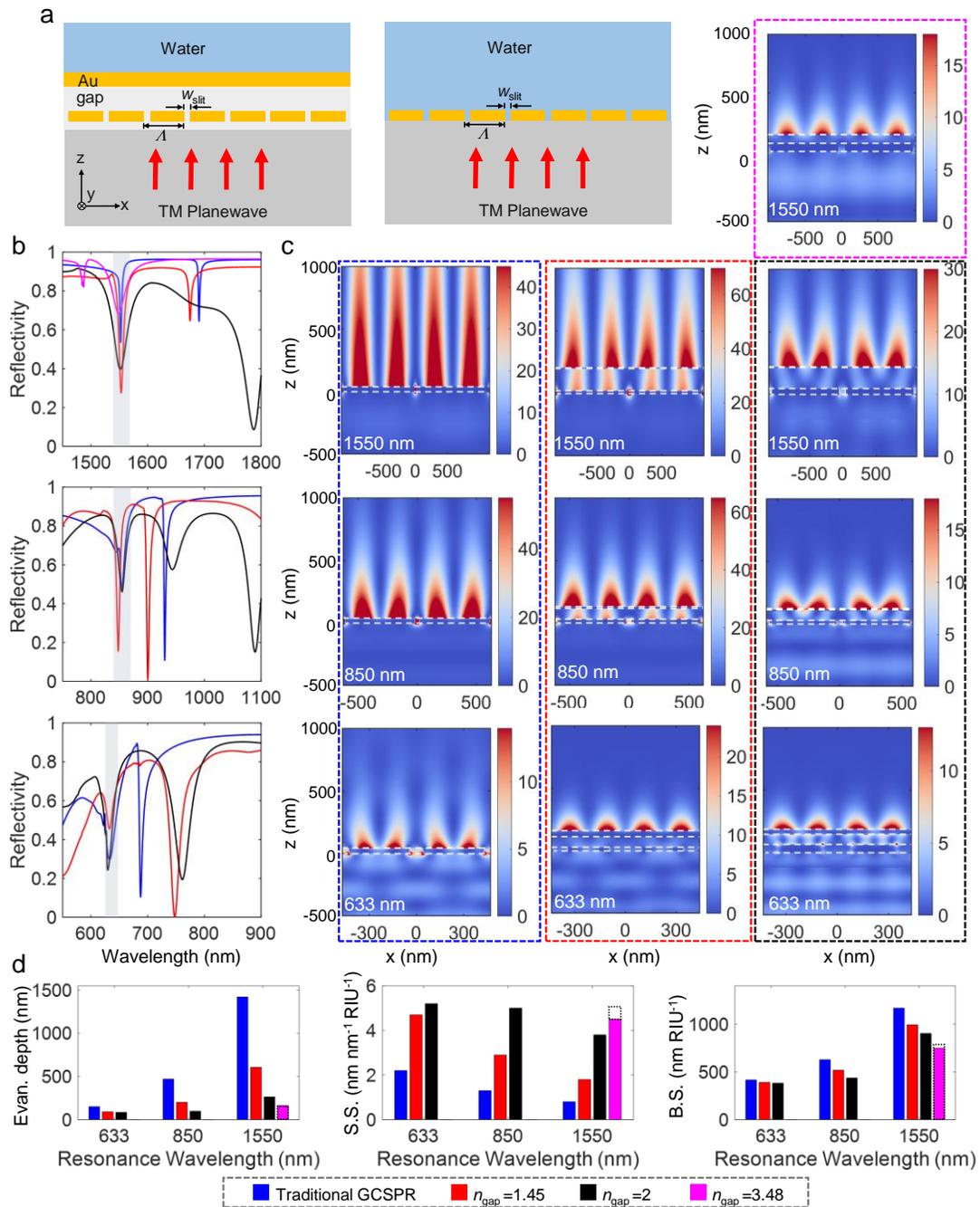

**Fig. 6. Nanoslit-grating-coupled meta-films for a wide range of wavelengths.** (**a**) The simulation model (not to scale), including both the meta-film and the traditional GCSPR. (**b**) Reflectivity spectra with 633 nm, 850 nm and 1550 nm resonance wavelengths, using the geometric parameters shown in Table I. (**c**) $|E|^2$ profiles at the resonance wavelengths, normalized to the incident light. (**d**) The evanescent depths, surface sensitivities and bulk sensitivities at different wavelengths, employing different gap refractive indices. The blue curves, frameworks and bars correspond to traditional GCSPR, while red, black and magenta correspond to meta-films with $n_{gap}$=1.45, 2.0 and 3.48, respectively.   Note that for the case of 1550 nm and $n_{gap}$=3.48, the surface sensitivity has been compromised to obtain a reasonably deep spectral resonance dip as we list the geometric parameters in Table I, the same of which has been plotted here. At the same time, the maximum surface sensitivity case for 1550 nm and $n_{gap}$=3.48 is shown by the dashed bars which overlay the magenta ones.

The same device structure as in the last section is used (Fig. 6a). $n_{gap}$ is set to 1.45 (SiO$_2$), 2.0 (Si$_3$N$_4$) and 3.48 (Si at 1550 nm), corresponding to different gap dielectric materials, respectively. Then, the geometric parameters of the devices are optimized to maximize the surface sensitivities at each target wavelength, using the particle swarm method (Table I). For comparison purpose, the traditional GCSPR structures are also calculated, in which the meta-film is replaced by a single-layer Au film embedded with a periodic nanoslit array (Fig. 6a), with their geometric parameters listed in Table 1 as well.

The calculated power reflectivity spectra are shown in Fig. 6b, with each sub-figure corresponding to one of the target wavelengths. The blue curves correspond to traditional GCSPR, and the red, black and magenta curves correspond to meta-film devices with $n_{gap}$ = 1.45, 2.0, and 3.48, respectively. In every situation, a resonance dip is clearly present showing efficient coupling between the device and the back illumination.

The calculated mode field profiles, $|E|^2$, are shown in Fig. 6c, following the same color identifications. As predicted earlier, in general, a higher $n_{gap}$ or a shorter wavelength corresponds to a shorter evanescent depth, $l_d$. The extracted $l_d$ values are plotted in Fig. 6d. In short, at 633 nm wavelength, $l_d$ drops from 149 nm for GCSPR to as short as 84 nm for the meta-film ($n_{gap}$=2.0), dropping by a factor of 1.8; at 850 nm, it drops from 470 nm for GCSPR to 97 nm ($n_{gap}$=2.0), by a factor of 4.9; at 1550 nm, it drops the most significantly, from 1416 nm for GCSPR to 164 nm ($n_{gap}$=3.48), by a factor as much as 8.6. Note that the silicon gap has only been considered for 1550 nm, since silicon becomes absorptive below 1 μm optical wavelength.

The surface sensitivities are also plotted in Fig. 6d. It shows that, at 850 nm, the optimized surface sensitivity reaches 5.0 nm nm$^{-1}$ RIU$^{-1}$, which is close to our earlier prediction assuming a lossless Au material and an infinitely thick Au substrate, the latter giving 5.1 nm nm$^{-1}$ RIU$^{-1}$. The results show that, the meta-film surface sensitivities are higher than those of GCSPR by a factor of 2.4, 3.9 and 5.6, at 633nm ($n_{gap}$=2.0), 850 nm ($n_{gap}$=2.0), and 1550 nm ($n_{gap}$=3.48), respectively. Particularly, the surface sensitivity at 1550 nm is not only over twice as high as that of GCSPR at 633 nm, but also nearly as high as those of the meta-films at 633 nm and 850 nm, making it a potentially compelling solution for plasmonic biosensing. In addition, the bulk sensitivities are plotted for reference, showing a trend of decreasing as $n_{gap}$ increases, as predicted earlier.

## VII. Discussion and Conclusion

A flat interface is crucial for achieving stable plasmonic biosensing performance, for which the SPP-MIM meta-film has proven itself to be a capable solution. In this work, we conduct a theoretical investigation of these meta-films, and show that they provide a strong combination of traditional SPR and LSPR. In brief, the hybridization process of *SPP+S* pushes its dispersion curve away from the water lightline, which can greatly reduce the evanescent depth in water, resulting in remarkably enhanced surface sensitivities. Meanwhile, the wide tunability of the MIM guided mode, *S*, makes it flexible to tune the *SPP+S* hybridized mode across a broad wavelength range, from visible to near infrared. By embedding a nanoslit grating in the meta-film, we preliminarily explore the performance limits of these grating-coupled sensors. Up to nearly six times enhancement of surface sensitivity at 1550 nm, compared with traditional GCSPR, is obtained. Significant enhancements are observed at 633 nm and 850 nm as well.

It is well known that LSPR devices are blessed with high surface sensitivities and wide tunabilities (e.g. the core-shell structures [36], [37]). The proposed meta-films possess these advantages as well. Together with an order of magnitude reduction in the evanescent depth (at 1550 nm), these meta-films show strong LSPR characteristics, yet maintaining a flat interface. We believe a whole family of new device designs will benefit from the SPP-MIM meta-film paradigm in the future, based on its high performance in various aspects. For example, it has already been employed for the successful demonstration of sensitive and stable biosensing at optical fiber tips [22]. Our work will provide a guide for the engineering of these devices.

Note that we used as thin as 2 nm of Au for some of the optimized device geometries (Table I). While this is difficult to achieve with traditional metal deposition techniques [38], recent advancements in nanotechnology show that it could be indeed possible [39], [40]. It is interesting to mention that, when the top metallic film becomes thin enough as a two-dimensional electron gas, the *H'* mode turns into so-called acoustic surface plasmons, which is an intensively studied subject for investigating the interactions between two-dimensional materials and plasmonics [41]. The abilities to manipulate molecular sensing or molecular spectroscopy by making an insulating layer into the conducting material, as in this or other work [42], presents intriguing opportunities for both optical material engineering and light-matter interactions. Further, if we incorporate more complicated material properties, such as birefringence, phonons and multiple layers, there could be many interesting discoveries ahead.

## Appendix
### Simulation Methods and Settings

Numerical simulations of the electromagnetic fields and the consequent dispersion curves, photonic band diagrams, reflection spectra and sensing performances were carried out using the finite-difference time-domain (FDTD) method, which has been performed with Ansys Lumerical FDTD. The perfectly matched layer (PML) boundary condition was applied along the *z*-direction, and the Bloch boundary condition was applied along the *x*-direction.

For the meta-films without the nanoslit gratings (Figs. 1-3), the dielectric constant of Au was taken to follow the Drude model [43], in which $\omega_P$ = 1.37×10$^{16}$ rad/s, and the ohmic loss was ignored. The simulation mesh had a grid size of 0.2 nm in the *z*-direction and 1 nm in the *x*-direction. To excite the guided modes in these structures, multiple randomly placed and oriented dipole sources were used, and Fourier transform of the time-varying fields was performed to obtain the modes.

For the nanoslit-grating-coupled devices (Figs. 5 and 6), the dielectric constant of Au was taken from ref. [43], which includes the ohmic loss. The grid size of the simulation mesh was from 1 to 5 nm within and near the devices, except for being 0.2 nm inside and near the top Au layer when its thickness was under 10 nm.

## DEFINITIONS OF SENSITIVITIES

The bulk sensitivity is defined as,

$$B.S. = \frac{d\lambda}{dn_B} \quad (1)$$

where $\lambda$ is the resonance wavelength of the sensor, and $n_B$ is the refractive index of the bulk background.

To define the surface sensitivity, in this work, a thin dielectric film is added to the surface of the sensor (the surface of the top Au film). A water background is assumed. The surface sensitivity is defined as,

$$S.S. = \frac{1}{h}\frac{\partial \lambda}{\partial n_S} \quad (2)$$

where $\lambda$ is the resonance wavelength of the sensor, $h$ is the thickness of the added film, and $n_S$ is the refractive index of the added film which is also called surface refractive index in this paper.

To calculate the grating-coupled sensitivities for a meta-film without a grating (Figs. 2-4), we let $\lambda$ in Eqs. (1) and (2) be the wavelength of the point on the dispersion curve (of the desired mode) at $k_x = 2\pi/\Lambda$, where $\Lambda$ is an imaginary grating period.

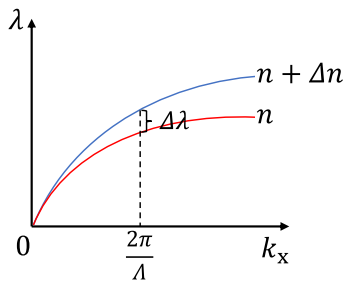

**Fig. 7.** Definition of grating-coupled sensitivity for a meta-film without a grating.